\documentclass[preprint2]{aastex}
\usepackage{graphicx}
\usepackage{epstopdf}

\shorttitle{ULXs and metallicity}
\shortauthors{Mapelli}

\begin{document}


\title{The missing link between ultraluminous X-ray sources and metallicity}


\author{Michela Mapelli}
\affil{INAF-Osservatorio Astronomico di Padova, Vicolo dell'Osservatorio 5, I--35122, Padova, Italy \email{michela.mapelli@oapd.inaf.it}}






\begin{abstract}
The nature of ultraluminous X-ray sources (ULXs) is still debated. Recent studies show that metal-poor massive stars can collapse into massive stellar black holes (MSBHs), that is black holes with mass $> 25$ M$_\odot$. Such MSBHs are sufficiently massive to explain most ULXs without requiring substantial violations of the Eddington limit. The recent finding of an anti-correlation between metallicity of the environment and number of ULXs per galaxy supports this hypothesis. We present the results of recent $N-$body simulations, including metallicity dependent stellar evolution, and we discuss the main pathways to produce X-ray binaries powered by MSBHs.
\end{abstract}





\section{Introduction}
Ultraluminous X-ray sources (ULXs) are point-like off-nuclear sources with X-ray luminosity, assumed isotropic, $L_{\rm X}>10^{39}$ erg s$^{-1}$. This corresponds to the Eddington limit of a $>7$ M$_\odot{}$ black hole (BH). Various explanations have been proposed for the nature of ULXs.  ULXs could be associated
with high-mass X-ray binaries (HMXBs) powered by stellar-mass BHs with anisotropic X-ray emission (e.g. King et al. 2001) or with super-Eddington accretion rate/luminosity (e.g. Begelman 2002; 
 Poutanen et al. 2007). 
  Some ULXs (e.g. Farrell et al. 2009) could be even associated with binaries powered by intermediate-mass BHs (IMBHs), i.e. BHs with mass $10^2\le{} m_{\rm BH}/{\rm M}_\odot{}\le{}10^5$.
 Recent models (e.g., Mapelli, Colpi \&{} Zampieri 2009; Zampieri \&{} Roberts 2009) suggest that a fraction of ULXs may be powered by massive stellar BHs (MSBHs), i.e. BHs in the mass range $25\le{} m_{\rm BH}/{\rm M}_\odot{}\le{}80$. These MSBHs might form from the collapse of massive stars, provided that the metallicity of the progenitor star is sufficiently low (e.g. Belczynski et al. 2010, hereafter B10).
Finally, the objects classified as ULXs might actually be a mixed bag of different sources.

The ULXs match the correlation between X-ray luminosity and star formation rate (SFR) reported by various studies (Grimm, Gilfanov \&{} Sunyaev 2003; Ranalli, Comastri \&{} Setti 2003). 
 A number of studies suggest an (anti-)correlation between ULXs and metallicity, and propose that this may be connected with the influence of metallicity on the evolution of massive stars (Pakull \&{} Mirioni 2002; Zampieri et al. 2004; Soria et al. 2005; Swartz, Soria \&{} Tennant 2008; Mapelli et al. 2009; Zampieri \&{} Roberts 2009; Mapelli et al. 2010, hereafter M10; Mapelli et al. 2011, hereafter M11). 

In this paper, we discuss the observational hints for a connection between ULXs and metallicity (Section 2), and we present the results of $N-$body simulations studying the population of X-ray binaries at various metallicities (Section 3). In Section 4, we summarize our main conclusions.
\section{ULXs and metallicity}
\begin{figure}
\epsscale{0.8}
\plotone{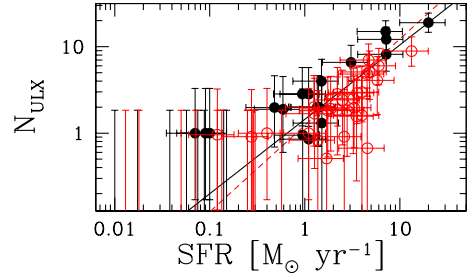}
\caption{N$_{\rm ULX}$ versus SFR in the sample by M11. Red open circles: host galaxies with $Z>0.2$ Z$_\odot{}$; black filled circles: host galaxies with $Z\le{}0.2$ Z$_\odot{}$. The solid black line is the power-law fit for the entire sample. The dashed red line is the power-law fit obtained assuming that the index of the power law is $=1$. The error bars on both the $x$ and the $y$ axis are 1 $\sigma{}$ errors.\label{fig1}}
\end{figure}
M11 analyze a sample of 66 late-type nearby galaxies. 
 64 galaxies were taken from the sample listed in Table~1 of M10. The remaining two are I~Zw~18 and the interacting pair SBS~0335052/SBS~0335052W1. These two objects are 
important, because they are the only galaxies with metallicity $Z < 0.03$ Z$_\odot{}$ and with published X-ray observations (Thuan et al. 2004). For each galaxy in the sample, M11 derive a fiducial value for the SFR, for the metallicity (adopting the calibration by Pilyugin \&{} Thuan 2005; when a metallicity gradient is available, M10 and M11 adopt the value of $Z$ at 0.7 Holmberg radii), and estimate the number of ULXs, N$_{\rm ULX}$, after subtracting the background contamination (see M10 for details). Fig.~\ref{fig1} shows N$_{\rm ULX}$ versus SFR for the galaxies in M11. This Figure points out that N$_{\rm ULX}$ scales almost linearly with the SFR, and it also shows that a galaxy with relatively low metallicity ($Z\le{}0.2$ Z$_\odot{}$, black filled circles) hosts on average more ULXs than a galaxy with the same SFR but higher metallicity ($Z>0.2$ Z$_\odot{}$, red open circles). 

This trend is confirmed in Fig.~\ref{fig2}, where N$_{\rm ULX}/{\rm SFR}$ versus $Z$ is shown for the same sample of galaxies. The red line in Fig.~\ref{fig2} comes from one of the models described in M10. To obtain the red line, we assumed that N$_{\rm ULX}=\epsilon{}$ N$_{\rm BH}$, where N$_{\rm BH}$ is the number of MSBHs per galaxy, while $\epsilon{}$ is the fraction of MSBHs that power ULXs. N$_{\rm BH}$ was estimated as described in eq.~1 of M10 (adopting the B10 model for BH formation and a Kroupa initial mass function), and we assume  $\epsilon{}=3.8\times{}10^{-4}$, which matches the data of M11.  
\section{Model and simulations}
\begin{figure}
\epsscale{0.8}
\plotone{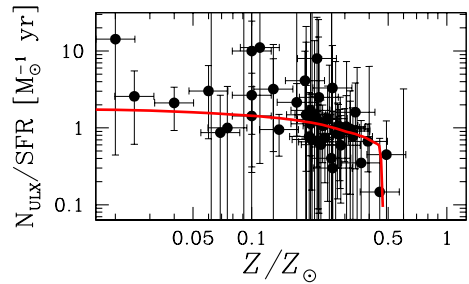}
\caption{Filled black circles: $N_{\rm ULX}/{\rm SFR}$ versus metallicity $Z$ in the sample by M11.  The error bars on both the $x$ and the $y$ axis are 1 $\sigma{}$ errors. The solid red line is the prediction by M10. See the text for more details. \label{fig2}}
\end{figure}

As shown in Fig.~\ref{fig2}, the model in which MSBHs power (a fraction of) ULXs is able to reproduce the metallicity trend in observed ULXs. On the other hand, Linden et al. (2010, hereafter L10) perform population synthesis simulations of binaries with different metallicities, and find that the contribution of MSBHs to bright X-ray binaries is marginal with respect to that of stellar-mass BHs emitting at super-Eddington rate. L10 explain the excess of ULXs in low-metallicity environments as the effect of metal-dependent stellar evolution, rather than as a consequence of the remnant mass. 

\begin{figure}
\epsscale{0.8}
\plotone{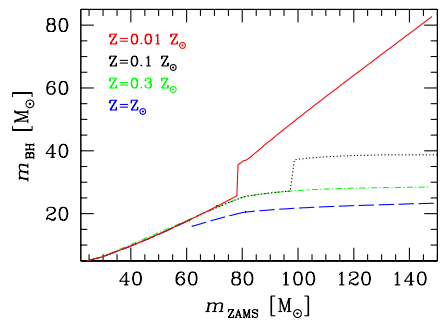}
\caption{Mass of the BH versus zero-age main sequence (ZAMS) mass of the progenitor star. Red solid line:  $Z=0.01$ Z$_\odot{}$; black dotted line:  $Z=0.1$ Z$_\odot{}$; green dot-dashed line: $Z=0.3$ Z$_\odot{}$; blue dashed line: $Z=1$ Z$_\odot{}$.\label{fig3}}
\end{figure}

L10 analysis considers only field binaries, which evolve in isolation. On the other hand, most stars, and especially massive stars, are expected to form in star clusters (SCs, e.g. Lada \&{} Lada 2003). Furthermore, a large fraction of ULXs are associated with OB associations and with young intermediate-mass SCs (e.g., Goad et al. 2002;  Zezas et al. 2002; Liu, Bregman \&{} Seitzer 2004; Soria et al. 2005; 
Abolmasov et al. 2007; 
Swartz, Tennant \&{} Soria 2009; Tao et al. 2011; Gris\'e et al. 2012).

SCs are often collisional environments, i.e. systems whose two-body relaxation time is shorter than their lifetime. This implies that most binaries in SCs do not evolve unperturbed, but undergo a number of three-body encounters (i.e. close encounters with single stars, e.g. Sigurdsson \&{} Phinney 1993). Three-body encounters can affect dramatically the orbital properties of binaries. Furthermore, dynamical exchanges can change the members of a binary. Thus, dynamics can 
alter completely the evolution of a binary with respect to the predictions by population synthesis codes.   

To check the combined influence of dynamics and metal-dependent stellar evolution on the population of X-ray sources in SCs, we ran a set of $N-$body simulations with the code Starlab (Portegies Zwart et al. 2001). 
We included recipes for metal-dependent stellar evolution and for the formation of MSBHs (see Mapelli et al. 2013) in the public version of the code. Fig.~\ref{fig3} shows the mass of the BH as a function of the zero-age main sequence (ZAMS) mass of the progenitor star, as implemented in our simulations.

 We ran over 300 simulations of young ($<100$ Myr) intermediate-mass ($3000-4000$ M$_\odot{}$) SCs, with metallicity $Z=0.01$, 0.1 and 1 Z$_\odot$.
The simulated SCs follow a King profile (with adimensional central potential $W_0=5$ and core radius $r_{\rm c}=0.4$ pc), adopt a Kroupa initial mass function, and an initial binary fraction $f_{\rm PB}=0.1$. The core collapse time for these systems is $\sim{}3$ Myr, short with respect to the lifetime of massive stars.

\begin{figure*}
\epsscale{0.7}
\plotone{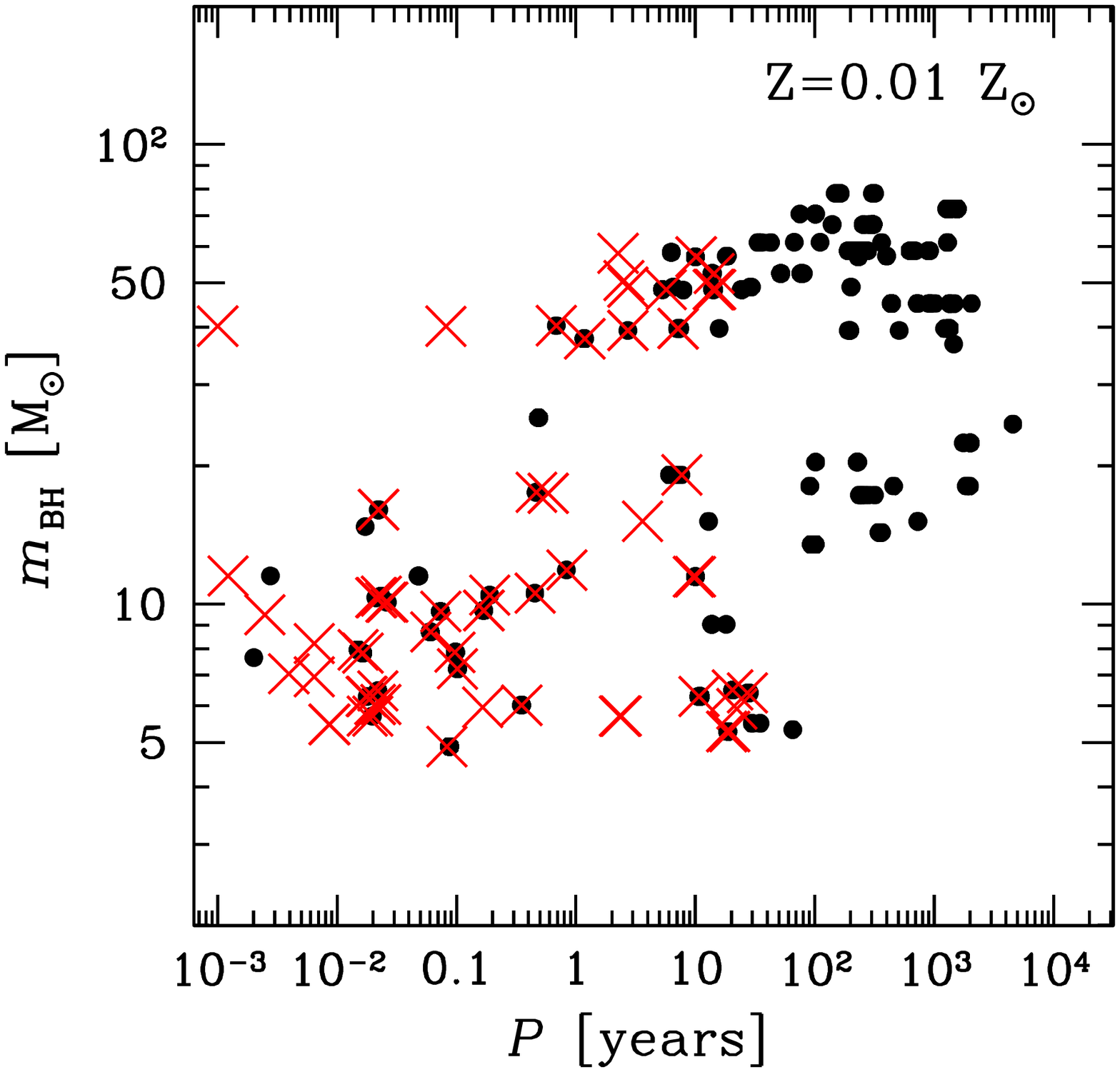}
\plotone{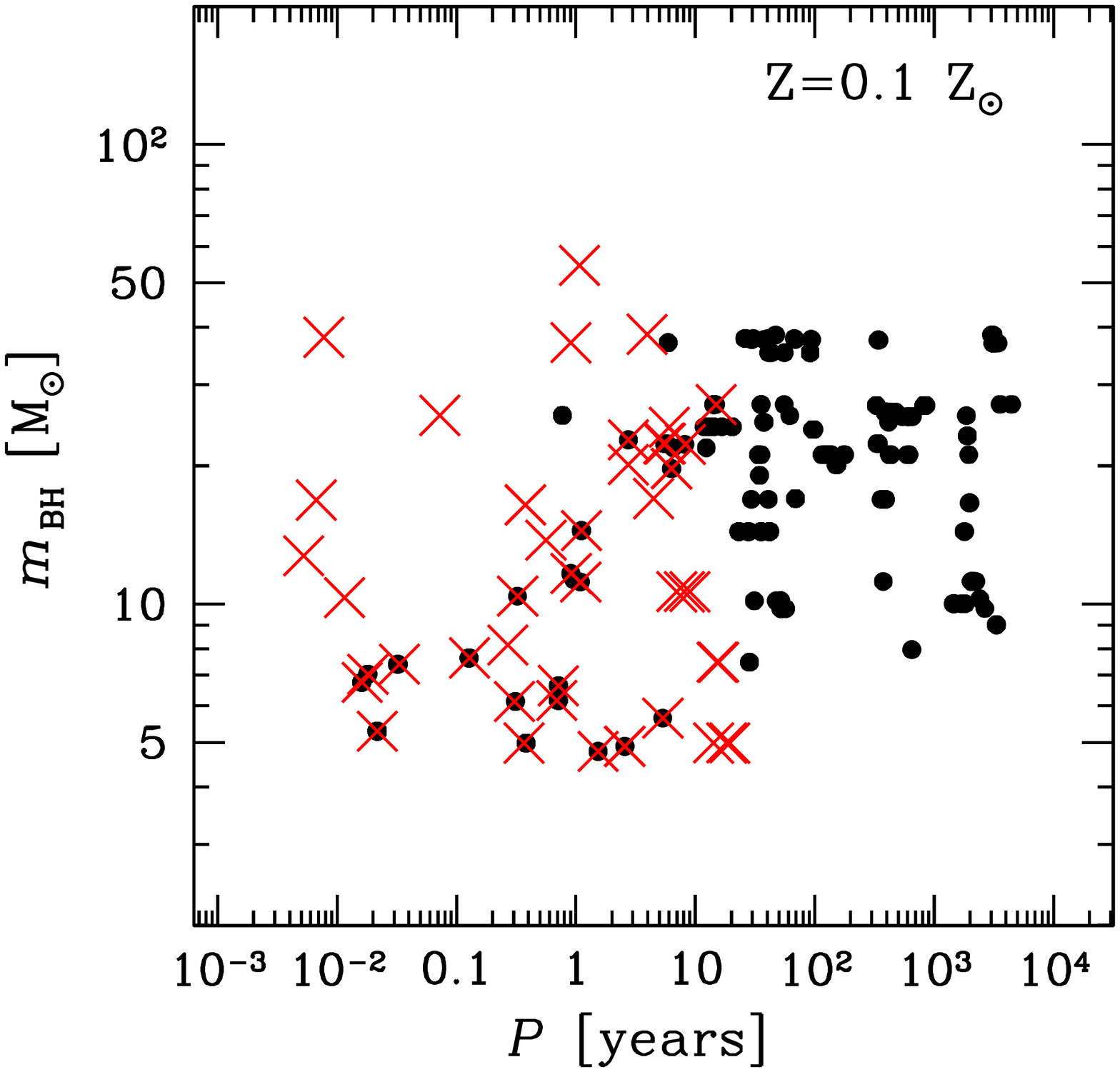}
\plotone{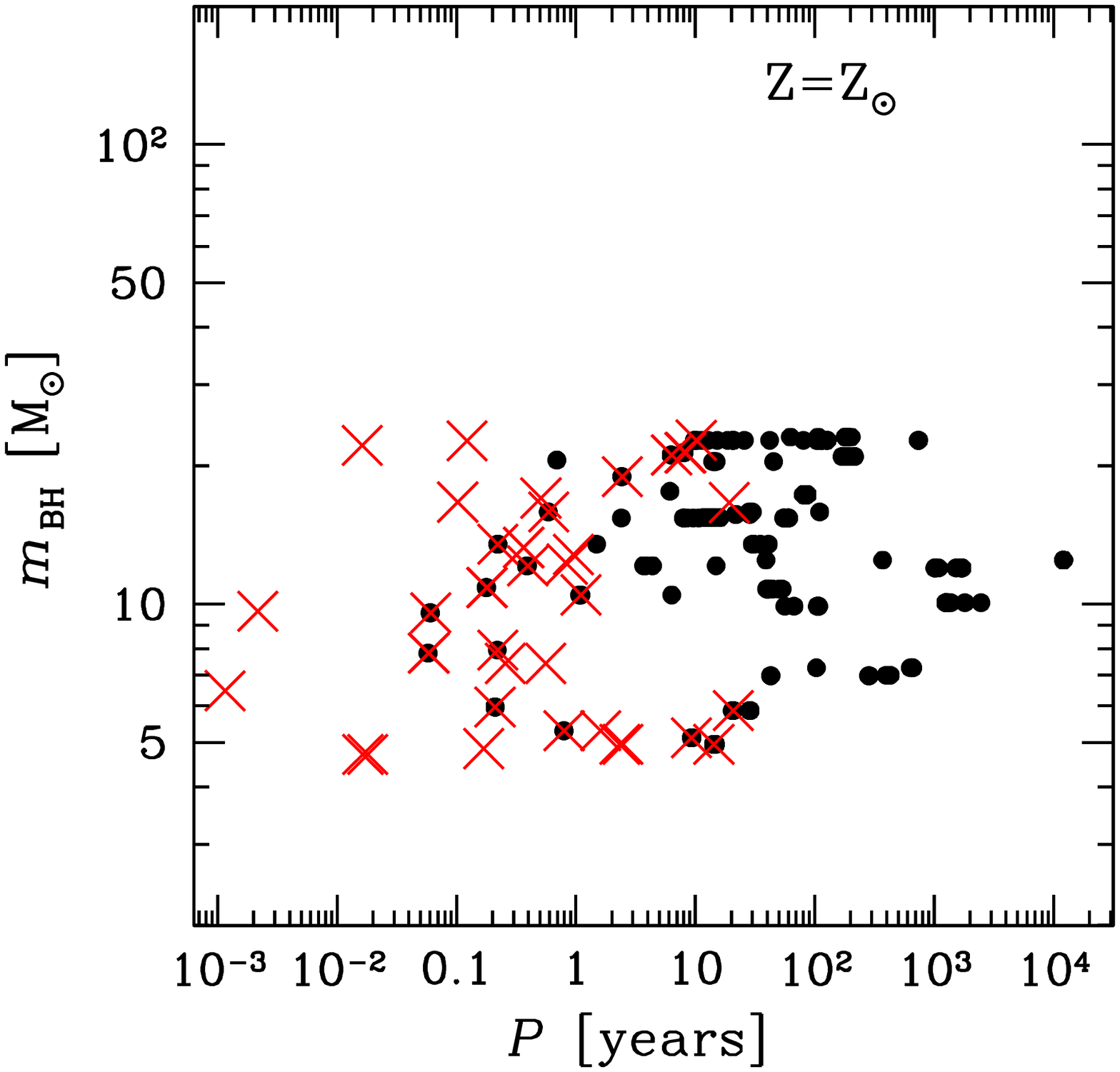}
\caption{Mass of the BH versus orbital period in the simulated BH binaries. Filled circles: wind-accretion systems; red crosses: RLO systems (at the first RLO epoch). From left to right:  0.01 Z$_\odot{}$,  0.1 Z$_\odot{}$, 1  Z$_\odot{}$. Each system can be identified by more than one point, when the period evolves significantly, as consequence of accretion, circularization, or dynamical interactions.\label{fig4}}
\end{figure*}

The simulations provide information about the formation and evolution of X-ray binaries, powered by both Roche lobe overflow (RLO) and wind accretion. Fig.~\ref{fig4} shows the mass of the BH versus the orbital period of RLO (crosses) and wind-accreting (circles) binaries. It is apparent that accreting binaries at low metallicity ($Z=0.01-0.1$ Z$_\odot$) can be powered by MSBHs: $10-20$ per cent of all MSBHs power wind-accreting systems, while  $5-10$ per cent of all MSBHs power RLO systems. Interestingly, the vast majority ($>90$ per cent) of accreting binaries powered by MSBHs underwent at least one dynamical exchange before starting the accretion. In most cases, the MSBH formed from a single star and then became member of a binary via exchange. 

\section{Conclusions}
In this paper, we reported the results of $N-$body simulations of intermediate-mass (3000-4000 M$_\odot{}$) young SCs with three different metallicities (Z=0.01, 0.1 and 1 Z$_\odot{}$), including metal-dependent stellar evolution recipes and binary evolution. The aim is to check the importance of dynamics for the population of X-ray sources at different metallicities, and to estimate the probability that a MSBH powers an X-ray binary.

We showed that $\sim{}5-10$ per cent of all the MSBHs formed at low metallicity ($Z=0.01-0.1$ Z$_\odot$) power RLO systems. The vast majority of accreting binaries powered by MSBHs underwent at least one dynamical exchange before starting the accretion. Our results are in agreement with L10, as the number of accreting MSBHs in unperturbed primordial binaries is negligible. The key result of our simulations is that MSBHs are efficient in powering X-ray binaries through dynamical evolution. This result indicates that MSBHs can power X-ray binaries in low-metallicity young SCs, and is very promising to explain the association of many ultraluminous X-ray sources with low-metallicity and actively star forming environments. 
\acknowledgments{\footnotesize{We made use of the public software package Starlab (version 4.4.4) and of the SAPPORO library to run Starlab on graphics processing units (GPUs). We acknowledge all the developers of Starlab, and especially its primary authors: Piet Hut, Steve McMillan, Jun Makino, and Simon Portegies Zwart. We thank the authors of SAPPORO, and in particular E. Gaburov, S. Harfst and S.  Portegies Zwart. We acknowledge the CINECA Award N.  HP10CXB7O8 and HP10C894X7, 2011 for the availability of high performance computing resources and support. MM acknowledges financial support from INAF through grant PRIN-2011-1.}}






\end{document}